\documentclass[12pt]{article}
\pdfoutput=1

\usepackage{amsmath,amssymb,amscd}
\usepackage{listings}
\usepackage{caption}
\usepackage{dsfont}
\usepackage{slashed}
\usepackage{color}
\usepackage{ulem}

\usepackage[pdftex]{graphicx}
\usepackage{epstopdf}
\usepackage{subfigure}
\usepackage{epsfig}
\usepackage{listings}
\usepackage{caption}
\usepackage{cite}

\usepackage{multirow}

\newcommand{\beq}{\begin{equation}}
\newcommand{\eeq}{\end{equation}}
\newcommand{\nbea}{\begin{align*}}
\newcommand{\neea}{\end{align*}}
\newcommand{\nbeq}{\begin{equation*}}
\newcommand{\neeq}{\end{equation*}}

 \usepackage{multirow}
\usepackage{array}
\newcolumntype{M}[1]{>{\centering\arraybackslash}m{#1}}
\newcolumntype{N}{@{}m{0pt}@{}}

\begin{document}


\pagestyle{empty}

\baselineskip=21pt
\rightline{{\fontsize{0.40cm}{5.5cm}\selectfont{KCL-PH-TH/2017-01, CERN-TH/2017-007}}}
\vskip 0.3in

\begin{center}

{\large {\bf FRB 121102 Casts New Light on the Photon Mass}}

\vskip 0.3in


{\bf Luca Bonetti}\textsuperscript{a,b,c},~
{\bf John Ellis}\textsuperscript{d,e},~
{\bf Nikolaos E. Mavromatos}\textsuperscript{d,e},~\\
{\bf Alexander S. Sakharov}\textsuperscript{f,g,h},~
{\bf Edward K. Sarkisyan-Grinbaum}\textsuperscript{h,i},~\\
{\bf Alessandro D.A.M. Spallicci}\textsuperscript{a,b,c}

\vskip 0.2in

{\small {\it

\textsuperscript{a}{\mbox Observatoire des Sciences de l'Univers en r\'egion Centre, UMS 3116, Universit\'e d'Orl\'eans}\\
\mbox{1A rue de la F\'erollerie, 45071 Orl\'eans, France}\\  
\vspace{0.25cm}
\textsuperscript{b}{\mbox P\^ole de Physique, Collegium Sciences et Techniques, Universit\'e d'Orl\'eans}\\
\mbox{Rue de Chartres, 45100 Orl\'eans, France}\\  
\vspace{0.25cm}
\textsuperscript{c}Laboratoire de Physique et Chimie de l'Environnement et de l'Espace, UMR 7328\\
Centre Nationale de la Recherche Scientifique\\
\mbox{LPC2E, Campus CNRS, 3A Avenue de la Recherche Scientifique, 45071 Orl\'eans, France}\\
\vspace{0.25cm}
\textsuperscript{d}Theoretical Particle Physics and Cosmology Group, Physics Department \\
King's College London, Strand, London WC2R 2LS, United Kingdom\\
\vspace{0.25cm}
\textsuperscript{e}Theoretical Physics Department, CERN, CH-1211 Gen\`eve 23, Switzerland\\
\vspace{0.25cm}
\textsuperscript{f}Department of Physics, New York University\\
4 Washington Place, New York, NY 10003, United States of America\\
\vspace{0.25cm}
\textsuperscript{g}Physics Department, Manhattan College\\
{\mbox 4513 Manhattan College Parkway, Riverdale, NY 10471, United States of America}\\
\vspace{0.25cm}
\textsuperscript{h}Experimental Physics Department, CERN, CH-1211 Gen\`eve 23, Switzerland \\
\vspace{0.25cm}
\textsuperscript{i}Department of Physics, The University of Texas at Arlington\\
{\mbox 502 Yates Street, Box 19059, Arlington, TX 76019, United States of America} \\}}

\vskip 0.2in

{\bf Abstract}

\end{center}

\baselineskip=18pt \noindent


{\small
The photon mass, $m_\gamma$, can in principle be constrained using measurements of the 
dispersion measures (DMs) of fast radio bursts (FRBs),
once the FRB redshifts are known. The DM of the repeating FRB 121102
is known to $< 1$\%, a host galaxy has now been identified with high confidence,
and its redshift, $z$, has now been determined with high accuracy: $z = 0.19273(8)$. Taking into account the
plasma contributions to the DM from the Intergalactic medium (IGM) and the Milky Way,
we use the data on FRB 121102 to derive the constraint
$m_\gamma \lesssim 2.2 \times 10^{-14}$~eV c$^{-2}$ ($3.9 \times 10^{-50}$~kg).
Since the plasma and photon mass contributions to DMs have different redshift dependences,
they could in principle be distinguished by measurements of more FRB redshifts, enabling the sensitivity 
to $m_\gamma$ to be improved.
}




\newpage
\pagestyle{plain}

The photon is generally expected to be massless, but a number of theorists have challenged this assumption,
starting from de Broglie and nowadays considering models with massive photons for dark energy and dark matter. Examples of mechanisms for
providing mass include Standard Model Extensions with supersymmetry and Lorentz invariance breaking~\cite{BDHS}
and Higgs mechanisms~\cite{ADG}. In view of these possibilities and its fundamental importance, it is important
to constrain the magnitude of the photon mass as robustly as possible. The most robust limits available are those
from laboratory experiments~\cite{wifahi71} - see \cite{OtherReviews,goni10} for reviews -
but these are much weaker than those derived from astrophysical observations. 
The Particle Data Group (PDG)~\cite{PDG} cites the upper limit $m_\gamma < 8.4 \times 10^{-19}$~eV c$^{-2}$ 
($= 1.5 \times 10^{-54}$~kg) \cite{ry97} obtained by modelling the magnetic field of the solar system~\cite{ry97,ry07}.
However, this limit relies on assumptions about the form of the magnetic field and does not discuss
measurement accuracy and errors. Another limit ($m_\gamma < 4 \times 10^{-52}$~kg) has been derived from atmospheric 
radio waves has been reported in~\cite{Fullekrug}. A more conservative approach was followed in an analysis of Cluster 
data~\cite{respva2016}, leading to an upper limit between $7.9 \times 10^{-14}$ and $1.9 \times 10^{-15}$~eV c$^{-2}$
($1.4 \times 10^{-49}$ and $3.4 \times 10^{-51}$ kg). It is clearly desirable to explore
more direct and robust astrophysical constraints on a possible photon mass.

This was the motivation for a study we made~\cite{BEMSSS} 
(see also~\cite{Meszaros}) showing how data from fast radio bursts (FRBs)
could be used to constrain $m_\gamma$. These have durations in the millisecond range, and
their signals are known to arrive with a frequency-dependent dispersion in time of the
$1/\nu^2$ form. This is the dependence expected from plasma effects, but a similar dispersion 
$\propto m_\gamma^2/\nu^2$ could also arise from a photon mass. The dispersions induced
by plasma effects and $m_\gamma$ both increase with distance (redshift $z$), 
but with different dependences on $z$. We note in this
connection that the lower frequencies of FRB emissions give a distinct advantage over
gamma-ray bursters and other sources of high-energy $\gamma$ rays for constraining
$m_\gamma$, since mass effects are suppressed for higher-energy photons~\footnote{In contrast,
sources of high-energy photons are better suited for probing models of Lorentz violation~\cite{LV}.}. Moreover,
using FRB emissions to constrain $m_\gamma$ is much more direct and involves fewer
uncertainties than using the properties of astrophysical magnetic fields~\footnote{For an early consideration of possible astrophysical photon
propagation delays, see~\cite{deBroglie}. For pioneering studies using astrophysical sources, see~\cite{flarestars} (flare stars) and~\cite{Crab} (Crab nebula), and
for an analogous subsequent study with greater sensitivity to the photon mass, see~\cite{Schaefer} (GRB 980703, $m_\gamma < 4.2 \times 10^{-47}$~kg). The most recent such
studies are those in~\cite{GRB050416A} (GRB 050416A, $m_\gamma < 1.1 \times 10^{-47}$~kg)
and~\cite{RadioMagellan} (radio pulsars in the Magellanic clouds, $m_\gamma < 2 \times 10^{-48}$~kg). Our limit on $m_\gamma$ is significantly stronger.}.

That said, although the large dispersion measures (DMs) and other arguments led to the
general belief that FRBs occur at cosmological distances, until recently no FRB redshift
had been measured. The first claim to measure a redshift was made for FRB 150418~\cite{FRB},
and this was the example we used in~\cite{BEMSSS} to show how FRB measurements
could in principle be used to constrain $m_\gamma$. However, the identification of the
host galaxy of FRB 150418 has subsequently been challenged~\cite{AGN}, and is now generally not
accepted~\cite{FRB121102}.

Our interest in the possibility of using FRBs to constrain $m_\gamma$ has recently been
revived, however, by the observation of repeated emissions from FRB 121102~\cite{FRB121102}. These
have permitted precise localisation of its host galaxy, which has made possible
a precise determination of its redshift, $z = 0.19273(8)$~\cite{FRB121102redshift}. 
This redshift determination
makes it possible, in turn, to use data on FRB 121102 to provide a robust constraint on $m_\gamma$,
as we discuss in this paper.

The dispersion measure (DM) is related to the frequency-dependent time lag of an FRB by
\begin{equation}
\Delta t_{\rm DM} \; = \; 415\left( \frac{\nu}{1\, {\rm GHz}}\right)^{-2} \; 
\frac{\rm DM}{\rm 10^5\; pc\; cm^{-3}}\; {\rm s}\, .
\label{DeltatDM}
\end{equation}
In the absence of a photon mass, the time-lag of an FRB is given by integrating the column density $n_e$ of free 
electrons along the line of flight of its radio signal
\begin{equation}
\Delta t_{\rm DM} \; = \; \int\frac{dl}{c}\frac{\nu_p^2}{2\nu^2} \, ,
\label{Deltatne}
\end{equation}
where $\nu_p=(n_ee^2/\pi m_e)^{1/2}=8.98\cdot 10^3n_e^{1/2}$~Hz.
Several sources contribute to this integrated column density of free electrons, notably
the Milky Way galaxy, the intergalactic medium (IGM) and the host galaxy.
The contribution to the DM~(\ref{DeltatDM}) of an FRB at redshift $z$ from the IGM
is given by the cosmological density fraction ${\rm\Omega_{IGM}}$ of 
ionized baryons~\cite{IGM,IGM2}:
\begin{equation}
{\rm DM_{IGM}}\; =\; \frac{3cH_0\Omega_{\rm IGM}}{8\pi Gm_p}H_e(z)\, ,
\label{D_M}
\end{equation}
where $H_0 = 67.8 (9)$~km/s/Mpc~\cite{PDG} is the present Hubble expansion rate~\footnote{We discuss later the 
impact of assuming a broader range $H_0 = 70 (4)$~km/s/Mpc~\cite{Living}.}, $G$ is the Newton constant, $m_p$ is the proton mass, and
the redshift-dependent factor
\begin{equation}
H_e (z) \; \equiv \; \int_0^z \frac{(1 + z^\prime) d z^\prime}{\sqrt{\Omega_\Lambda + (1 + z^\prime)^3 \Omega_m}} \, ,
\label{He}
\end{equation}
where $\Omega_\Lambda = 0.692 (12)$ and $\Omega_m = 0.308 (12)$~\cite{PDG}.
For comparison, the difference in time lags between photons of energies $E_{1,2}$ due to a non-zero photon mass has the form:
\begin{equation}
\Delta t_{m_\gamma} = \frac{m_{\gamma}^2}{2 H_0} \cdot \left(\frac{1}{E_1^2}-\frac{1}{E_2^2}\right) \cdot H_{\gamma}(z)\, ,
\label{arrivalT1}
\end{equation}
where we use natural units $h = c = 1$, and~\cite{stretch1,stretch2}
\begin{equation}
H_{\gamma} (z) \; \equiv \; \int_0^z \frac{d z^\prime}{(1 + z^\prime)^2\sqrt{\Omega_\Lambda + (1 + z^\prime)^3 \Omega_m}} \, .
\label{Hg}
\end{equation}
As already commented, the time lags due to the IGM and a possible photon mass have different dependences 
(\ref{He}, \ref{Hg}) on the redshift.
The uncertainties in the cosmological parameters and the measurement of the redshift measurement of FRB 121102
are taken into account in our analysis, with the uncertainties in the former being much more important, as we see later.

\begin{figure}
\centering
\includegraphics[scale=0.7]{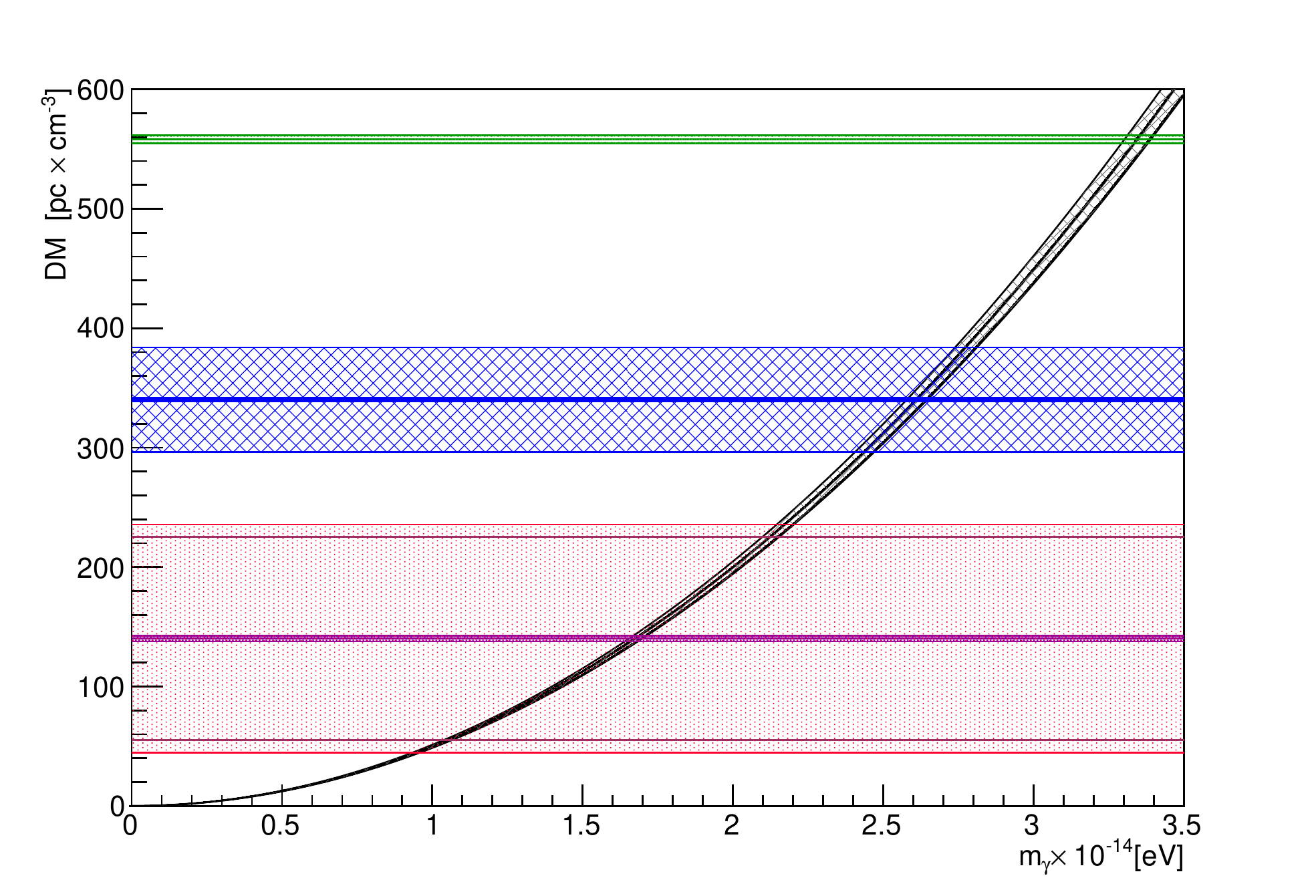}
\caption{\it Contributions to the dispersion measure (DM) budget for FRB 121102.
The top (green) band represents the experimental measurement of the total DM $= 558.1 \pm 3.3$~pc cm$^{-3}$.
The middle (blue) band shows the extragalactic contribution, as obtained by subtracting
from the central value of the total DM the galactic contribution
DM$_{\rm MW} \equiv$ DM$_{\rm NE2001} + $DM$_{\rm halo} = 218$~pc cm$^{-3}$~\protect\cite{FRB121102redshift},
to which is assigned an estimated uncertainty of 20\%. The bottom (pink) band is obtained by subtracting also the estimated
contribution from the intergalactic medium (IGM):
DM$_{\rm IGM} \simeq 200$~pc cm$^{-3}$~\protect\cite{FRB121102redshift}, 
which has an uncertainty of $85$~pc cm$^{-3}$ associated with inhomogeneities
in the IGM, and a smaller uncertainty associated with the
cosmological parameters (indicated by the narrow magenta band).
The outer pink band shows the total uncertainty in the residual DM after subtraction of DM$_{\rm MW}$ and DM$_{\rm IGM}$,
with errors added in quadrature.
The curved (black) band shows the possible contribution of a non-zero photon mass,
$m_\gamma$, also including the uncertainties in cosmological parameters and the redshift
of FRB 121102.
}
\label{fig:DMplot}
\end{figure}

The top (green) band in Fig.~\ref{fig:DMplot} shows
the total DM $= 558.1 \pm 3.3$~pc cm$^{-3}$ measured for FRB 121102~\cite{FRB121102}.
The most conservative
approach to constraining $m_\gamma$ would be to set to zero the other contributions,
and assign this total DM to
a possible photon mass. However, this is surely over conservative, and a reasonable
approach is to subtract from the total DM the expected contribution from the
Milky Way~\cite{FRB121102}, DM$_{\rm MW}$, which is the sum of contributions from
the disk~\cite{NE2001}: DM$_{\rm NE2001} \simeq 188$~pc cm$^{-3}$ and the halo:
DM$_{\rm halo} \simeq 30$~pc cm$^{-3}$~\cite{FRB121102redshift}, to which we assign an overall uncertainty
of 20\%, namely $44$~pc cm$^{-3}$~\cite{FRB121102redshift}, leaving the middle (blue) band in
Fig.~\ref{fig:DMplot} that is centred at $340$~pc cm$^{-3}$. From this we may also subtract the contribution from the IGM, which
is estimated within the $\Lambda$CDM model to be $\simeq 200$~pc cm$^{-3}$~\cite{IGM,IGM2,FRB121102redshift}. To this is assigned
an uncertainty of $85$~pc cm$^{-3}$ associated with inhomogeneities in the IGM~\cite{McQuinn,FRB121102redshift}, which is
much larger than the 1.2\% variation associated with uncertainties in the cosmological parameters
(shown as the narrow magenta band). The bottom (pink) band in
Fig.~\ref{fig:DMplot}, centred at $140$~pc cm$^{-3}$, shows the effect of subtracting these
contributions from the measured DM for FRB 121102.

After subtracting these contributions, we are left with a residual DM $= 140$~pc cm$^{-3}$
with a total uncertainty of $\pm 96$~pc cm$^{-3}$, shown as the outer pink band,
where the error is calculated by combining in quadrature the uncertainties in the experimental 
measurement of the total DM, the uncertainty in DM$_{\rm MW}$, and the uncertainties in DM$_{\rm IGM}$
associated with the  cosmological parameters $H_0, \Omega_\Lambda$ and $\Omega_m$ and possible inhomogeneities. 
One cannot exclude the possibility that all the residual DM of FRB 121102 is due to the host galaxy,
which is estimated to lie within the range $55 \lesssim$ DM$_{\rm Host} \lesssim 225$~pc cm$^{-3}$~\cite{FRB121102redshift}. 
However, in the absence of detailed information about the host galaxy, when constraining the photon mass
we allow conservatively for the possibility that all the residual DM is due to $m_\gamma \ne 0$.

The curved band in Fig.~\ref{fig:DMplot} shows the possible contribution to the DM of FRB 121102 of a
photon mass, as a function of $m_\gamma$: DM $= 10^5 m_{\gamma}^2H_{\gamma}/(415 A^2 h_0)$,
where $H_\gamma$ is given in (\ref{Hg}), $A=1.05\cdot 10^{-14}$ ev s$^{-1/2}$ and $h_0 \equiv H_0/100$km/s/Mpc. 
The width of this band is due to the uncertainties in the
cosmological parameters $H_0, \Omega_\Lambda$ and $\Omega_m$~\cite{PDG}, and the uncertainty in the determination of the redshift
of FRB 121102. 
Assuming that the photon mass contribution to the total 
DM of FRB 121102 lies within the range allowed for the residual DM,
after subtraction of the Milky Way and IGM contributions and taking their uncertainties into account,
we find $m_\gamma \lesssim 2.2 \times 10^{-14}$~eV c$^{-2}$ ($3.9 \times 10^{-50}$~kg)~\footnote{This limit
would increase to $m_\gamma \lesssim 2.3 \times 10^{-14}$~eV c$^{-2}$ ($4.1 \times 10^{-50}$~kg) if the more
relaxed range $H_0 = 70 (4)$~km/s/Mpc~\cite{Living} were used for $H_0$. On the other hand, it would
decrease to $m_\gamma \lesssim 1.8 \times 10^{-14}$~eV c$^{-2}$ ($3.2 \times 10^{-50}$~kg) if the minimum
estimate of DM$_{\rm Host}$~\cite{FRB121102redshift} was taken into account.}. This limit is similar to, though
slightly weaker than, that obtained from similar considerations of FRB 150418~\cite{BEMSSS,Meszaros}, 
whose redshift is now contested, as discussed earlier~\cite{AGN}. If FRB 150418
was indeed at a cosmological distance, using its DM value determined in~\cite{FRB} and the same values of
$H_e (z)$ and $H_{\gamma} (z)$ as in the 
present analysis, we find that the inferred limits on $m_\gamma$ would coincide if FRB 150418 had a redshift $z=0.38$, instead of 
the value $z=0.492$ reported in~\cite{FRB} and challenged in~\cite{AGN}.

How could this constraint be improved in the future? Clearly it is desirable to reduce the uncertainties
in the modelling of the Milky Way and IGM contributions. Also, the limit could be strengthened by a
redshift measurement for an FRB at higher $z$, if the uncertainty in the IGM contribution can be controlled.
Finally, as remarked in~\cite{BEMSSS}, comparing the DMs for FRBs with different redshifts could
enable the IGM and $m_\gamma$ contributions to be disentangled, in view of their different
dependences (\ref{He}, \ref{Hg}) on $z$. 

A hitherto unexplored window at very low frequencies in the MHz-KHz region could be opened by a space mission consisting
of a swarm of nanosatellites~\cite{swarm}. One possible configuration would be orbiting the Moon, where it would be
sufficiently away from the ionosphere to avoid terrestrial interference, and would have stable conditions for calibration during observations. 
Such low frequencies would offer a sensitive probe of any delays due to a non-zero photon mass.

\section*{Acknowledgements}

The research of J.E. and N.E.M. was supported partly by the STFC Grant ST/L000326/1. The work of A.S.S.
was supported partly by the US National Science Foundation under Grants PHY-1505463
and PHY-1402964.

\end{document}